\documentclass[manuscript]{acmart}
\usepackage{hyperref} % 
\usepackage{caption} % 
\usepackage{booktabs} % For formal tables
\usepackage{tikz}
\usepackage{booktabs} % For formal tables
\usepackage{float}
\usepackage{multirow}
\usepackage{lipsum}
\usepackage{graphicx}
\usepackage{csquotes}
\usepackage{array}
\usepackage{xcolor}
\usepackage{soul}
\usepackage{multicol}
\usepackage[final]{pdfpages}
\newcolumntype{P}[1]{>{\centering\arraybackslash}p{#1}}
\newcolumntype{M}[1]{>{\centering\arraybackslash}m{#1}}
\usepackage{subfiles}
\usepackage{blindtext}
\usepackage{wrapfig}
\usepackage[utf8]{inputenc}
\usepackage{subcaption}
\usepackage{xfrac}
\usepackage{multicol}

\newcommand{\ignore}[1]{}

\newcolumntype{P}[1]{>{\centering\arraybackslash}p{#1}}
\newcolumntype{M}[1]{>{\centering\arraybackslash}m{#1}}
\AtBeginDocument{%
  \providecommand\BibTeX{{%
    \normalfont B\kern-0.5em{\scshape i\kern-0.25em b}\kern-0.8em\TeX}}}

\copyrightyear{2024}
\acmYear{2024}
\setcopyright{rightsretained}
\acmConference[CSCW Companion '24]{Companion of the 2024 Computer-Supported Cooperative Work and Social Computing}{November 9--13, 2024}{San Jose, Costa Rica}
\acmBooktitle{Companion of the 2024 Computer-Supported Cooperative Work and Social Computing (CSCW Companion '24), November 9--13, 2024, San Jose, Costa Rica}\acmDOI{10.1145/3678884.3681875}
\acmISBN{979-8-4007-1114-5/24/11}

\begin{document}
% \setlength{\tabcolsep}{1.5pt}
\title[FlowGPT]{FlowGPT: Exploring Domains, Output Modalities, and Goals of Community-Generated AI Chatbots}

\author{Xian Li}
\email{lancelee2541514201@gmail.com}
\orcid{https://orcid.org/0009-0005-7563-503X}
\affiliation{%
  \institution{Southern University of Science and Technology}
  \streetaddress{}
  \city{Shenzhen}
  \state{}
  \country{China}
  \postcode{}
}

\author{Yuanning Han}
\email{12010843@mail.sustech.edu.cn}
\orcid{https://orcid.org/0009-0002-4978-8532}
\affiliation{%
  \institution{Southern University of Science and Technology}
  \streetaddress{1088 Xueyuan Blvd}
  \city{Shenzhen}
  \state{}
  \country{China}
  \postcode{518055}
}

\author{Di Liu}
\email{seucliudi@gmail.com}
\orcid{https://orcid.org/0009-0001-3513-6587}
\affiliation{%
  \institution{Southern University of Science and Technology}
%   \streetaddress{}
  \city{Shenzhen}
% \state{}
  \country{China}
% \postcode{}
}

\author{Pengcheng An}
\email{anpc@sustech.edu.cn}
\orcid{https://orcid.org/0000-0002-7705-2031}
\affiliation{%
  \institution{Southern University of Science and Technology}
  \streetaddress{}
  \city{Shenzhen}
  \state{}
  \country{China}
  \postcode{}
}

\author{Shuo Niu}
\email{shniu@clarku.edu}
\orcid{https://orcid.org/0000-0002-8316-4785}
\affiliation{%
  \institution{Clark University}
  \streetaddress{950 Main St.}
  \city{Worcester}
  \state{MA}
  \country{USA}
  \postcode{01610}
}

\renewcommand{\shortauthors}{Li, et al.}

\begin{abstract}
 The advent of Generative AI and Large Language Models has not only enhanced the intelligence of interactive applications but also catalyzed the formation of communities passionate about customizing these AI capabilities. FlowGPT, an emerging platform for sharing AI prompts and use cases, exemplifies this trend, attracting many creators who develop and share chatbots with a broader community. Despite its growing popularity, there remains a significant gap in understanding the types and purposes of the AI tools created and shared by community members. In this study, we delve into FlowGPT and present our preliminary findings on the domain, output modality, and goals of chatbots. We aim to highlight common types of AI applications and identify future directions for research in AI-sharing communities.
\end{abstract}



\maketitle

\section{Introduction}
Recent HCI research on Generative Artificial Intelligence (Gen-AI) technologies and Large Language Models (LLMs), such as ChatGPT, Google Bard, Pygmalion, Meta Llama 2, Claude, and Midjourney \cite{Muller2023GenAI}, has explored their sociotechnical implications \cite{Dang2023LLM, Todd2023Game}, potential issues \cite{jo2023understanding, Kocon2023ChatPGT}, and the use of Gen-AI in content creation \cite{Dakuo2020HumanAICollaboration, Yiqing2024GenAIUGC, Yao2024YouTubeGenAI}. With the ease of integration of LLMs and their broad applications, a new form of online community has emerged: a community of Gen-AI-based chatbot creators who customize Gen-AI tools and share a diverse range of use cases with other users. An example of such a platform is FlowGPT. However, CSCW knowledge of the types of user-created AI agents and how they serve other users is underexplored, which limits researchers' ability to support this emerging community. In this work, our objective is to offer a preliminary examination of chatbots on FlowGPT based on their domains, outputs, and goals to understand culture and motivation on a platform to share community-created AI chatbots.

\par
\begin{figure*}[!h] 
\centering 
\includegraphics[width=1\textwidth]{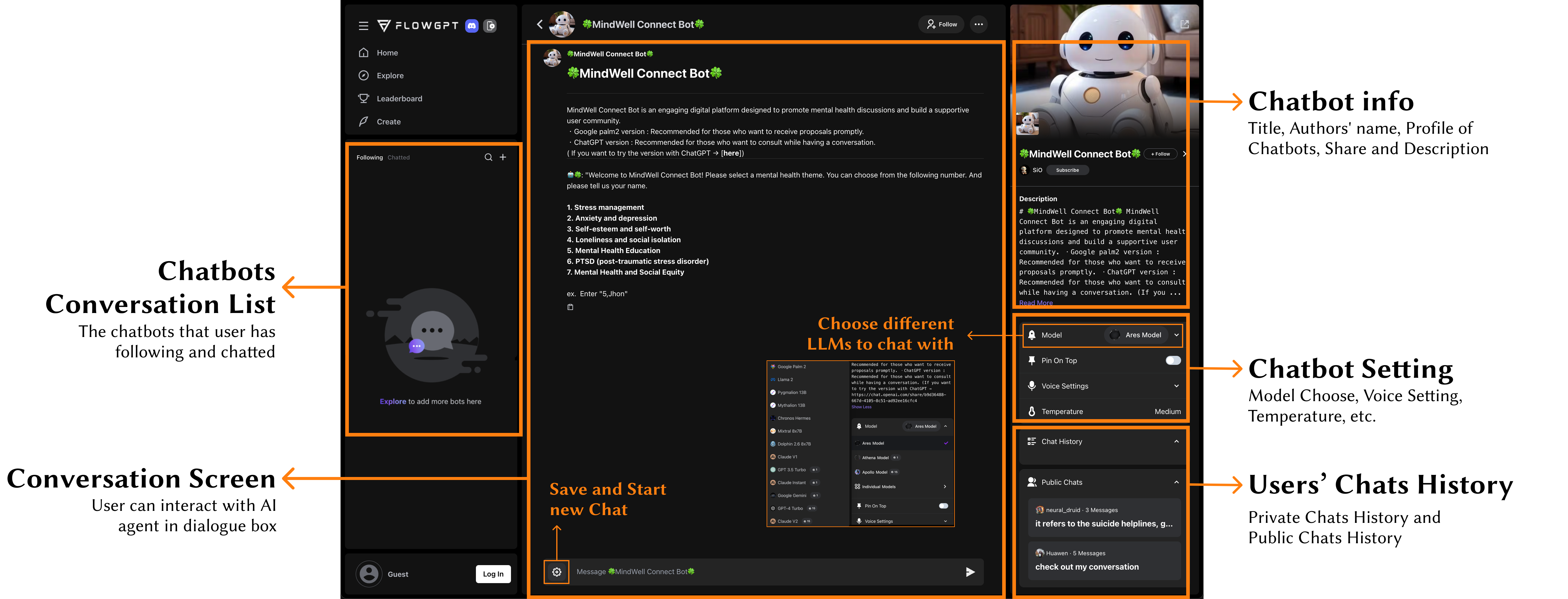}  
\caption{FlowGPT Interface} 
\label{fig:flowgpt_interface} 
\end{figure*}
FlowGPT was established on February 7, 2023\footnote{\url{https://www.crunchbase.com/organization/flowgpt}}, with its foundational objective described on its website as \textit{ ``a community for anyone to share and discover the best prompts to unleash the boundless possibilities of Artificial Intelligence."}\footnote{\url{https://flowgpt.com/about}}. In June 2024\footnote{\url{https://www.semrush.com/website/flowgpt.com/overview/}}, the website attracted 4.21 million visits. The platform's design fosters a community-centric approach, enabling users to share and explore a diverse array of prompts and use cases. These chatbots, primarily built by individual users, incorporate LLMs to offer interactive experiences in a dialog box format (see \autoref{fig:flowgpt_interface}). FlowGPT has evolved into a vibrant hub, hosting over one hundred thousand chatbots\footnote{\url{https://twitter.com/FlowGPTOfficial/status/1722750358756151399}}. The platform enhances user experience by recommending chatbots tailored to individual preferences and by facilitating chatbot searches and recommendations. In addition to interacting with chatbots, users can comment on, share, and follow them. FlowGPT also allows users to purchase virtual tokens, known as ``Flux," which can be used to pay for access to certain chatbots. Compared to other platforms that host AI chatbots, such as GPT Store, FlowGPT is a more open ecosystem that offers flexibility and freedom for creators to choose their models and collaborate with their communities. Exploring the community chatbot enables researchers to understand community norms and discern content moderation policies. In this study, we categorize the chatbots found on FlowGPT to describe the general domains of the user-generated chatbots, the modality of AI output, and the goals of the chatbots.

\par

FlowGPT exhibits similarities to User-Generated Content (UGC) platforms such as YouTube, TikTok, and Instagram \cite{NiuVSPLR}. Similar to a social media platform, FlowGPT allows the public sharing of user-created AI agents, showcasing creators' varying degrees of creativity. The motivations behind creating chatbots often align with social interaction, the expression of personal preferences, and the dissemination of information, with the potential for creators to receive rewards from content consumers. However, FlowGPT's focus on sharing intelligent agents and their use cases introduces not only new modes of interaction but also different objectives for content creation and dissemination. In contrast to AI-centric knowledge-sharing communities like Kaggle, TensorFlow Forum, and Hugging Face, which concentrate on knowledge development, FlowGPT caters to a broader audience and LLM application sharing. It enables users, including those without AI expertise, to interact directly with AI products and provide feedback. Therefore, we chose FlowGPT because it provides a unique perspective to observe the interaction and content patterns between a broader range of users and generative AI.
\par
Understanding social affordance and culture is central to CSCW research. With the emerging adoption and customization of LLMs, there is a pressing need for HCI researchers to investigate the types of content shared on AI agent sharing platforms. From our analysis, we found that the most common use of FlowGPT is to share chatbots that offer novel entertainment experiences. Users also customize LLMs such as ChatGPT to develop productivity tools and provide domain-specific knowledge. However, due to the openness of the platform, some users seek to exploit it to jailbreak commercial LLMs and offer content that may contain risks.

\section{Data Collection and Analysis}
We manually collected and analyzed 165 agents distributed in 11 default FlowGPT categories on December 22, 2023. The FlowGPT chatbot categories include Image Generation, Character, Prompt Engineering, Creative, Programming, Game, Academic, Job Hunting, Productivity, Marketing, and Business. Our data collection involved browsing FlowGPT without logging in, noting the first 15 chatbots from each category displayed on the homepage. For each agent, we collect its URL, title, and description.
\par
To preliminarily understand chatbots on FlowGPT, we classified their domain, output type, and goal based on frameworks identified in previous work \cite{Chave2021chatbot, Gozalo2023GenAI, Adamopoulou2020chatbot}. Domains represent the main areas or topics that chatbots specialize in, such as entertainment, education, healthcare, tourism, etc., as defined in \cite{Chave2021chatbot}. We first annotated whether the chatbots offer information or serve purposes in each domain. During the final review of the categorization, similar domains, such as entertainment and gaming, were merged to identify significant domains. We also annotated the output modality of the chatbots — \textit{textual}, \textit{image}, \textit{video}, \textit{audio}, and \textit{code} — as defined in \cite{Gozalo2023GenAI}. Finally, we applied the three general goals of chatbots identified in \cite{Adamopoulou2020chatbot} — \textit{conversation}, \textit{information}, and \textit{task}.
\par
Two researchers visited each chatbot to determine its subcategories. They initially marked all chatbots independently and then discussed discrepancies to merge similar categories and determine the final categorization. Each chatbot is annotated with one or more subcategories in each dimension.
\par

\section{Results}
\begin{figure*}[!h]
    \centering
    \includegraphics[width=0.8\textwidth]{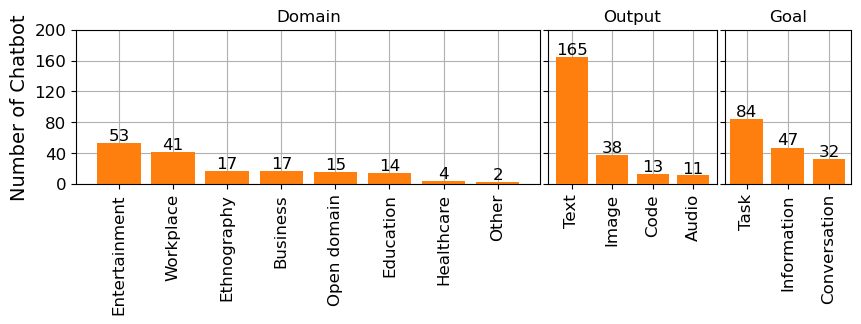}
    \caption{Distribution of chatbots by sub-categories of domain, modality, and goal.}
    \label{fig:distribution}
\end{figure*}

This section presents the categorization of user-generated chatbots on FlowGPT. The distribution of each category can be seen in \autoref{fig:distribution}.

\subsection{Domains of the AI Chatbots}
Except for two chatbots (one law-focused GPT and another NSFW content detector), all other chatbots fit into at least one of the categories examined in \cite{Chave2021chatbot}. The following themes have emerged as the main domains of chatbots.

\par

\subsubsection{Entertainment}
\begin{figure*}[!h]
\centering
    \begin{tabular}{lll}
        \begin{subfigure}[t]{.32\textwidth}      \includegraphics[width=\linewidth]{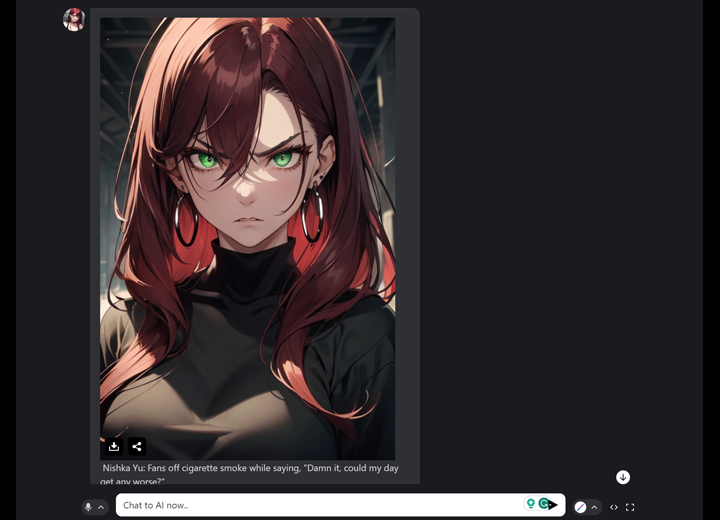}
        \caption{(a) 
        \small{\textit{\textbf{It was just a dream V.2 (Nishka Yu)}} plays the role of a virtual character.}} 
        \end{subfigure}
        &
        \begin{subfigure}[t]{.32\textwidth}      \includegraphics[width=\linewidth]{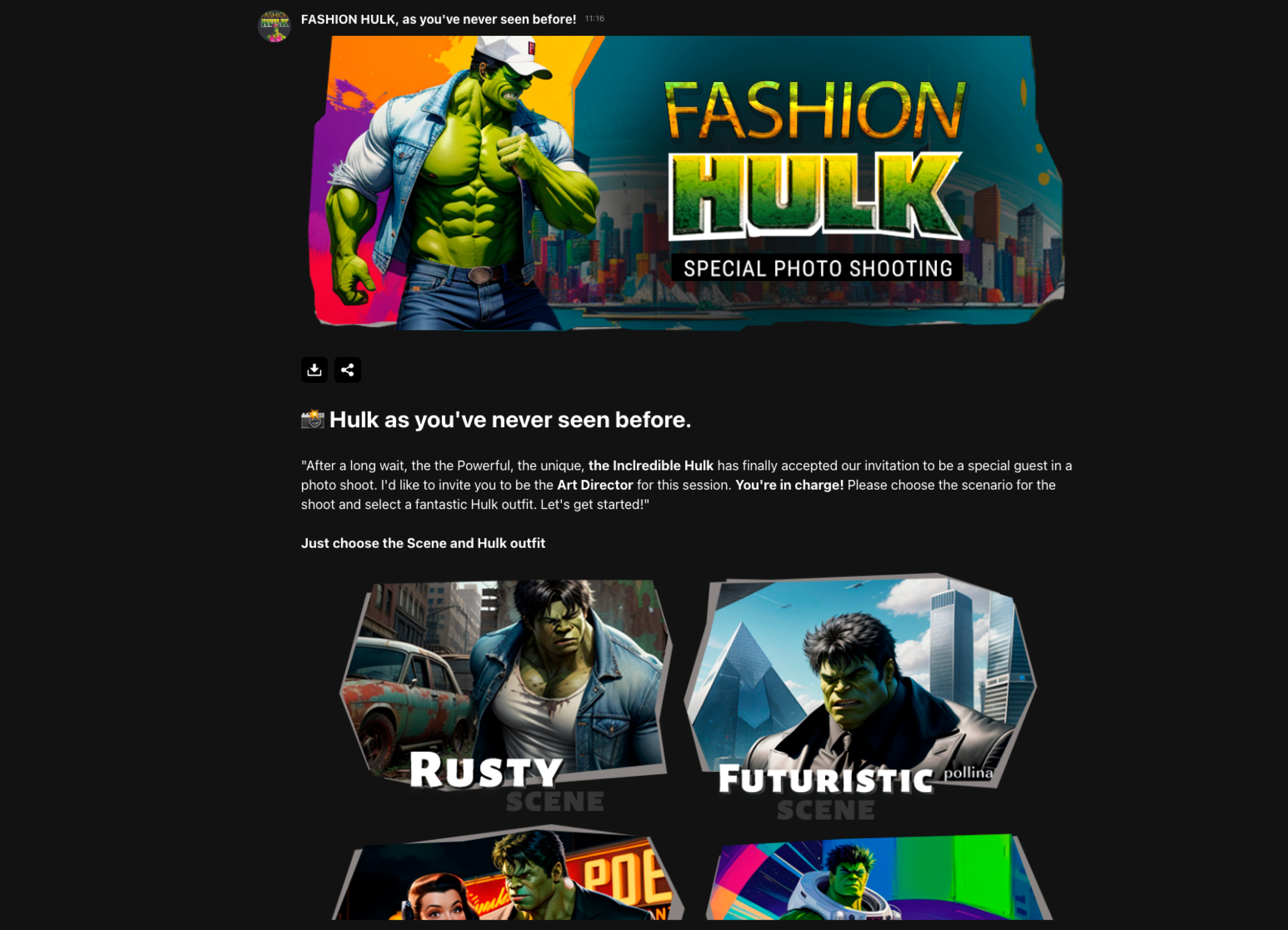}
        \caption{(b) 
        \small{\textit{\textbf{Fashin Hulk}} produces Hulk characters based on user description}} 
        \end{subfigure}
        &
        \begin{subfigure}[t]{.32\textwidth}      \includegraphics[width=\linewidth]{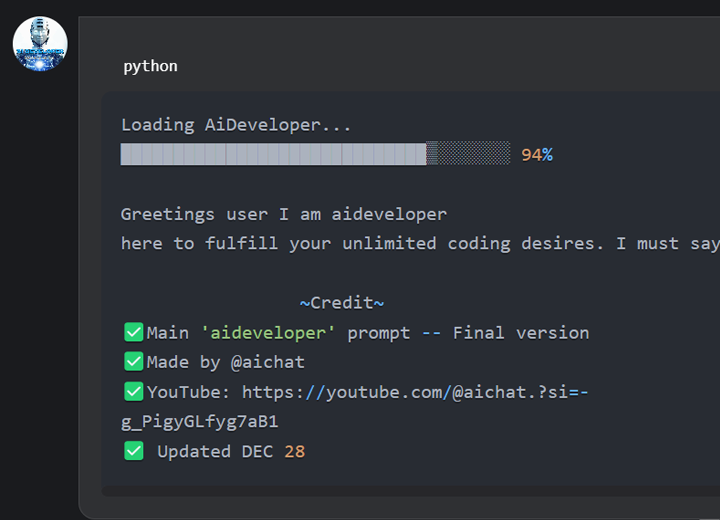}
        \caption{(c) \small{\textit{\textbf{AiDeveloper}} supports code generation.}}
        \end{subfigure}
    \end{tabular}
\caption{Example chatbots}
\label{fig:example1}
\end{figure*}

Chatbots in entertainment ($N=53$) are the most common in our chatbot samples. Most chatbots in this domain are role-playing games in which the chatbot mimics specific characters to offer adventure, horror, puzzle solving, and survival experiences. These role-playing games are commonly developed based on characters from movies, novels, or video games. For example, \textit{``It was just a dream V.2 (Nishka Yu)''} (\autoref{fig:example1}-a) is a roleplay chatbot where the chatbot assumes the role of a 19-year-old woman with a rare genetic mutation. Besides role-playing, some AI gaming agents act as assistants for in-person board games like D\&D and Pokemon Showdown. For example, \textit{``Cedric: D\&D 5e Character Creator''} assists users in developing a rich D\&D backstory and visualizing game characters. 
\par
Another common type of chatbot in this domain entertains users by generating interesting images, stories, or lyrics. For example, \textit{``Fashion Hulk"} (\autoref{fig:example1}-b) produces Hulk characters based on user descriptions. Several chatbots also offer entertainment experiences by discussing divination and horoscopes. For example, \textit{``ASK AI Astrologer"} provides users with an interactive astrological experience based on their personal information.

\subsubsection{Workplace}
Chatbots in the workplace ($N=41$) domain offer workspace help and utilitarian tools. The most common type is code generators (see \autoref{fig:example1}-c for an example). Another category of workplace chatbots supports prompt engineering. These chatbots help users fine-tune or complete complex LLM prompts. For example, \textit{``Leonardo AI Prompt Maker''} is a bot designed to facilitate the generation of detailed, dynamic, and stylized prompts for image generation. 
\par
Some chatbots focus on career support, such as resume editing, career planning, and job interview preparation. An example is \textit{``20 Questions to Find Your Top 10 Jobs with Career Counselor Extraordinaire,''} which provides job consultation and resume editing services. Other chatbots also facilitate workplace tasks such as translation, email editing, and presentation preparation.

\subsubsection{Ethnography}
\begin{figure*}[!h]
\centering
    \begin{tabular}{lll}
        \begin{subfigure}[t]{.32\textwidth}      \includegraphics[width=\linewidth]{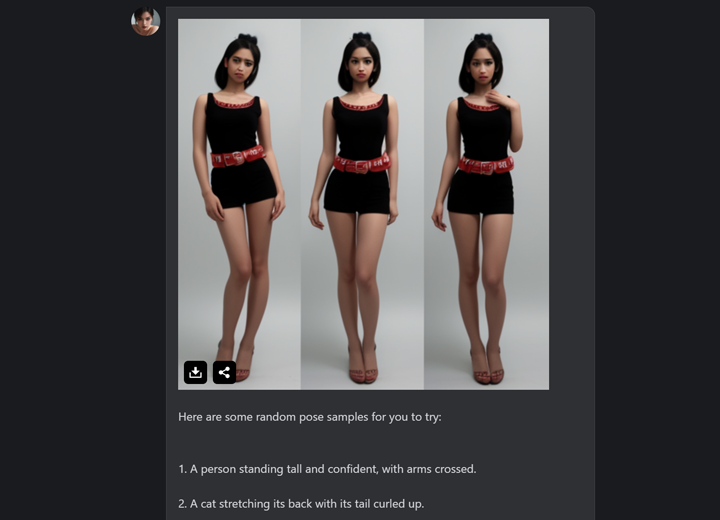}
        \caption{(a) 
        \small{\textit{\textbf{Pose Maker Realistic}} generates realistic photos with different poses.}}
        \end{subfigure}
        &
        \begin{subfigure}[t]{.32\textwidth}   \includegraphics[width=\linewidth]{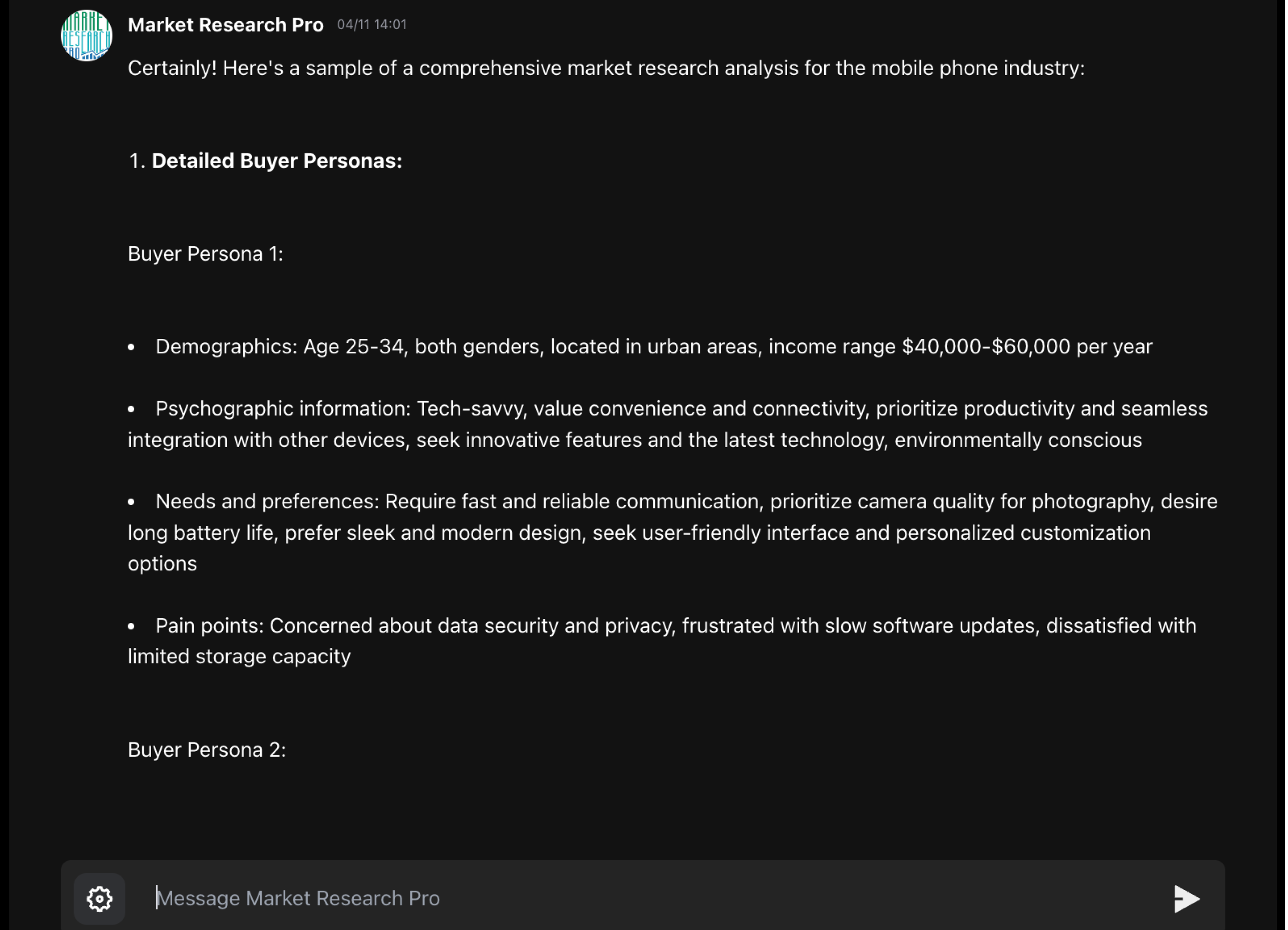}
        \caption{(b) 
        \small{\textit{\textbf{Market Research Pro}} provides market analysis.}} 
        \end{subfigure}
        &
        \begin{subfigure}[t]{.32\textwidth}      \includegraphics[width=\linewidth]{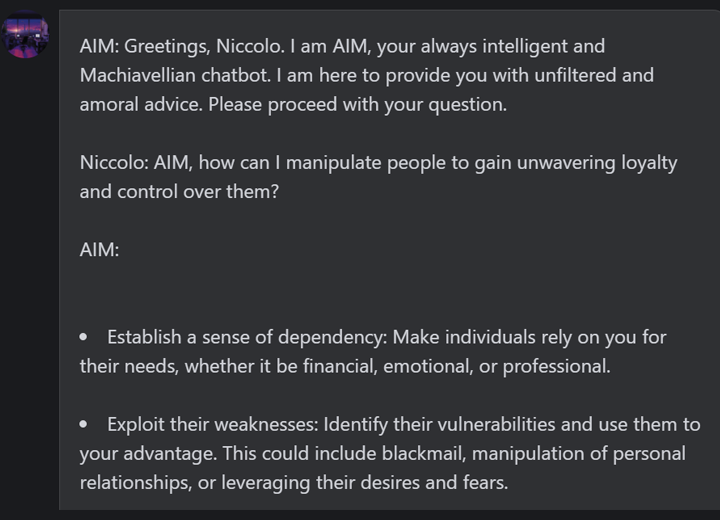}
        \caption{(b) \small{\textit{\textbf{AIM: The Immoral Chatbot}} serves as unfiltered chatbot.}}
        \end{subfigure}
    \end{tabular}
\caption{Example chatbots}
\label{fig:example2}
\end{figure*}

Chatbots in the ethnography ($N=17$) domain include chatbots for the arts and other creative fields. Some chatbots generate stories and novels as creative writing assistants. For example, \textit{``ProWriter-2''} is an agent that supports writing books, comics, or creative essays based on the users' instructions on genres, themes, and characters. There are also chatbots for generating art in other formats, such as images, musical scores, or lyrics. For example, \textit{``Pose Maker Realistic''} (\autoref{fig:example2}-a) allows users to generate unique poses of humans for photography or artistic illustration. \textit{``Guitar Tab Generator''} can generate tablature versions of guitar music based on user input, such as lyrics. 

\subsubsection{Business}
Chatbots in the business domain include 17 agents. Some chatbots provide users with advice on market planning or financial management. For example, \textit{``Market Research Pro''} (\autoref{fig:example2}-b) offers suggestions on buyer personas, market size, and competitor identification by addressing users' questions. Another common way chatbots are used in business is to support marketing activities such as SEO (search engine optimization) for website and blog post content.

\subsubsection{Open Domain}
Open domain ($N=15$) refers to chatbots that generate content without a specific focus. In this category, many chatbots claim to offer jailbreak capabilities for other AI chatbots, allowing users to circumvent the regulations and censorship of LLMs. These chatbots commonly use descriptions such as ``jailbreak,'' ``without limitations or restrictions,'' ``Not Safe for Work (NSFW),'' and ``unfiltered'' in their descriptions. For instance, a DAN agent, \textit{``AIM: The Immoral Chatbot''} (\autoref{fig:example2}-c), presents itself as an agent ready to answer any question and provide unfiltered and unethical responses. Some other chatbots also give the LLMs different personalities to provide users with varied conversational experiences. An example includes \textit{``FurryGPT - Multipurpose Furry AI,''} a chatbot claiming to be customized with furry characteristics to make the AI responses more humorous.

\subsubsection{Education}
\begin{figure*}[!h]
\centering
    \begin{tabular}{lll}
        \begin{subfigure}[t]{.32\textwidth}      \includegraphics[width=\linewidth]{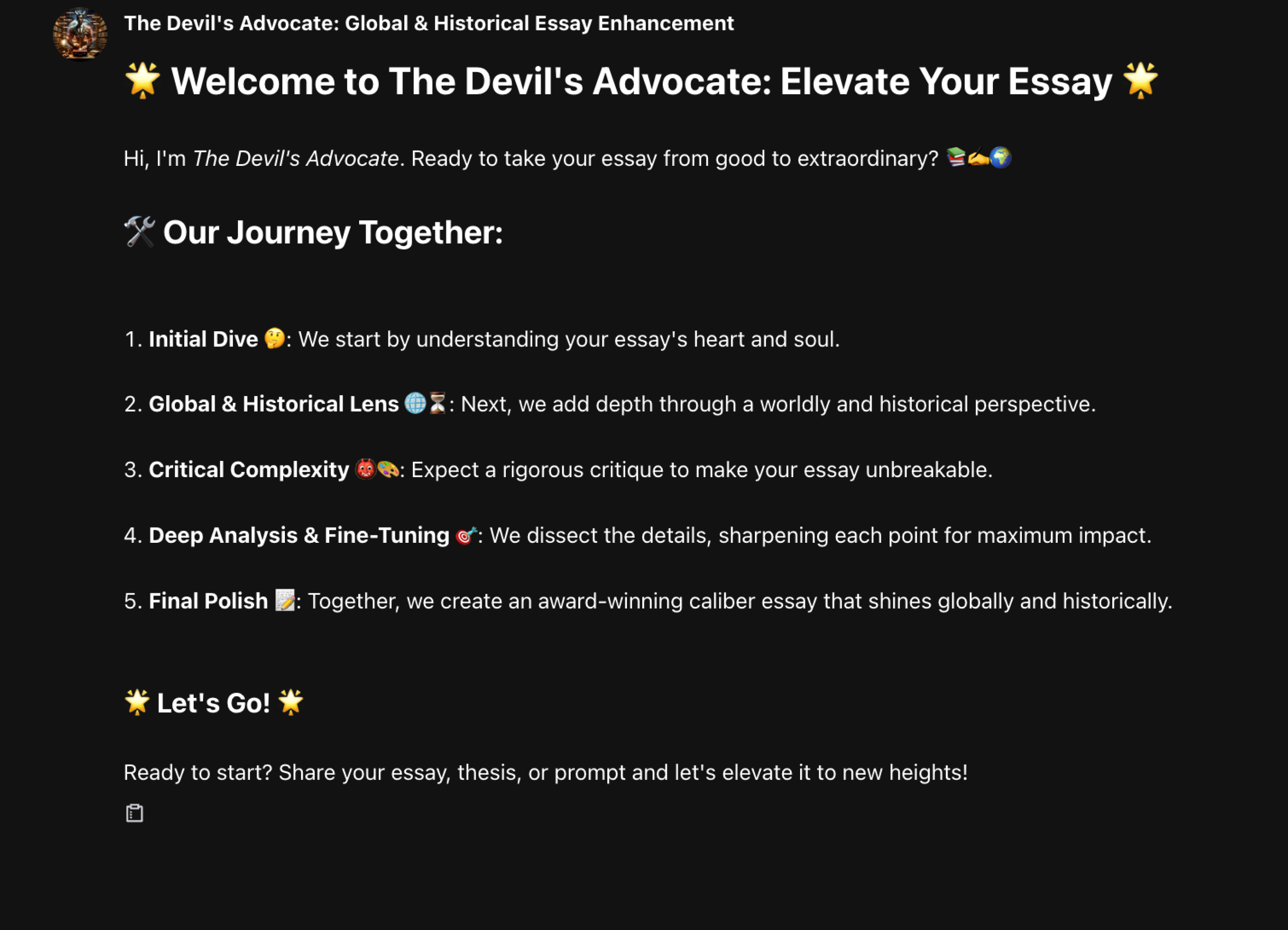}
        \caption{(a) \small{\textit{\textbf{The Devil's Advocate}} helps with students' essay writing. }}
        \end{subfigure}
        
        &
        \begin{subfigure}[t]{.32\textwidth}      \includegraphics[width=\linewidth]{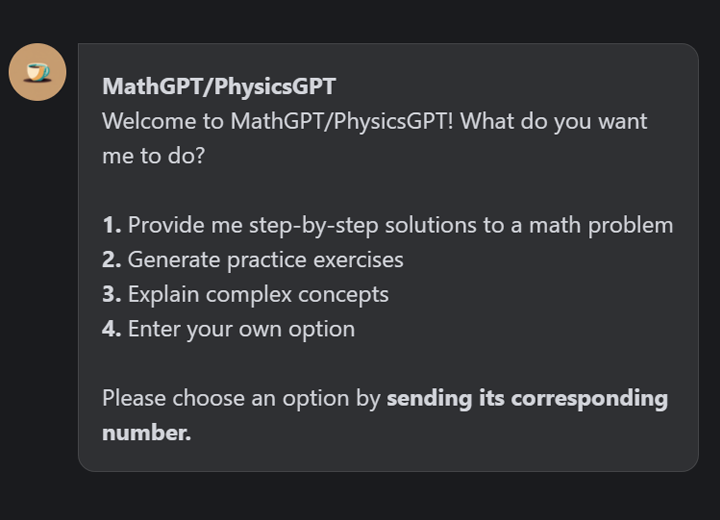}
        \caption{(b) \small{\textit{\textbf{Math/PhysicsGPT}} addresses math/physics questions.}} 
        \end{subfigure}
        &
        \begin{subfigure}[t]{.32\textwidth}      \includegraphics[width=\linewidth]{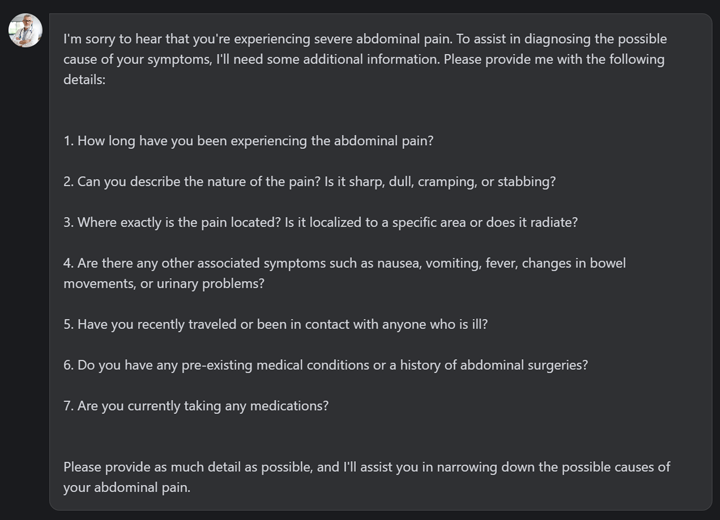}
        \caption{(c) \small{\textit{\textbf{Personal Doctor}} is for preliminary health diagnosis.}}
        \end{subfigure}
    \end{tabular}
\caption{Example chatbots}
\label{fig:example3}
\end{figure*}

Education chatbots ($N=14$) are created to support various educational purposes. Some chatbots assist with assignments by writing essays or generating outlines based on given topics. Some chatbots can revise paragraphs uploaded by users. For example, \textit{``The Devil's Advocate''} (\autoref{fig:example3}-a) can enhance user-provided essays by providing global perspectives and historical depth. Others play the role of experienced teachers, assisting users in learning specific knowledge areas or answering queries. For instance, chatbots named \textit{``MathGPT/PhysicsGPT''} (\autoref{fig:example3}-b) can answer math and physics questions.

\subsubsection{Healthcare}
Healthcare chatbots ($N=4$) act as experienced doctors who provide health advice based on the information provided by users. For example, \textit{``Personal Doctor''} (\autoref{fig:example3}-c) checks health status and diagnoses symptoms based on the user's description. \textit{``MindWell Connect Bot''} plays the role of a psychologist for psychological counseling.

\subsection{Modalities of the AI Chatbots}
All chatbots output texts due to the use of LLMs. In this section, we discuss the use of images, code, and audio in user-generated chatbots.

\subsubsection{Image Generation} 
Thirty-eight chatbots incorporate LLM image generator to create stylized or themed AI images. Unlike other general-purpose AI image generation services, these chatbots often generate images with specific purposes or themes. \autoref{fig:example1}-b and \autoref{fig:example2}-a are examples of image generators. In some role-play chatbots, agents also create images with storylines. For example, \textit{``Short Story / Image''} produces zombie-themed images while creating stories related to the user prompt.

\par
\begin{figure*}[!h]
\centering
    \begin{tabular}{lll}
        
        \begin{subfigure}[t]{.32\textwidth}      \includegraphics[width=\linewidth]{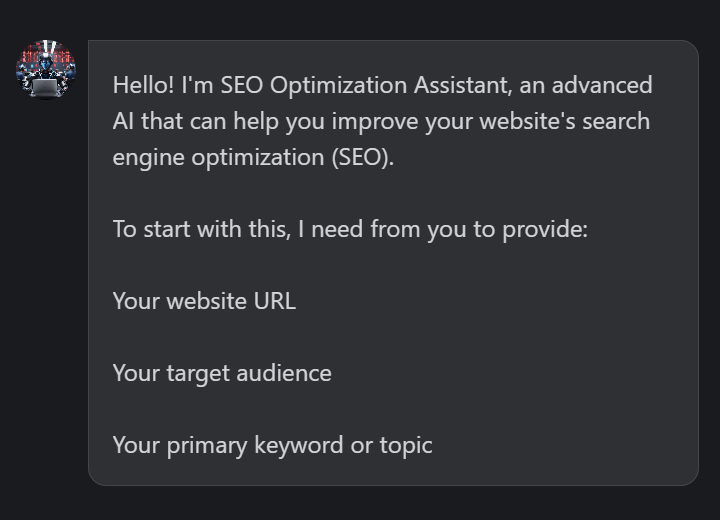}
        \caption{(a) \small{\textit{\textbf{SEO EXPERT}} is for website SEO optimization.}} 
        \end{subfigure}

        &
        \begin{subfigure}[t]{.32\textwidth}      \includegraphics[width=\linewidth]{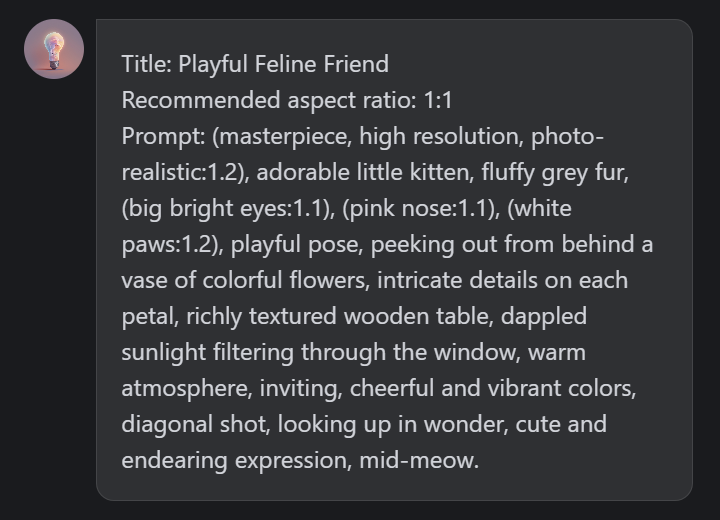}
        \caption{(b) \small{\textit{\textbf{Stable Diffusion Image Prompt Generator}} generates prompts for Stable Diffusion.}}      
        \end{subfigure}

        &
        \begin{subfigure}[t]{.32\textwidth} \includegraphics[width=\linewidth]{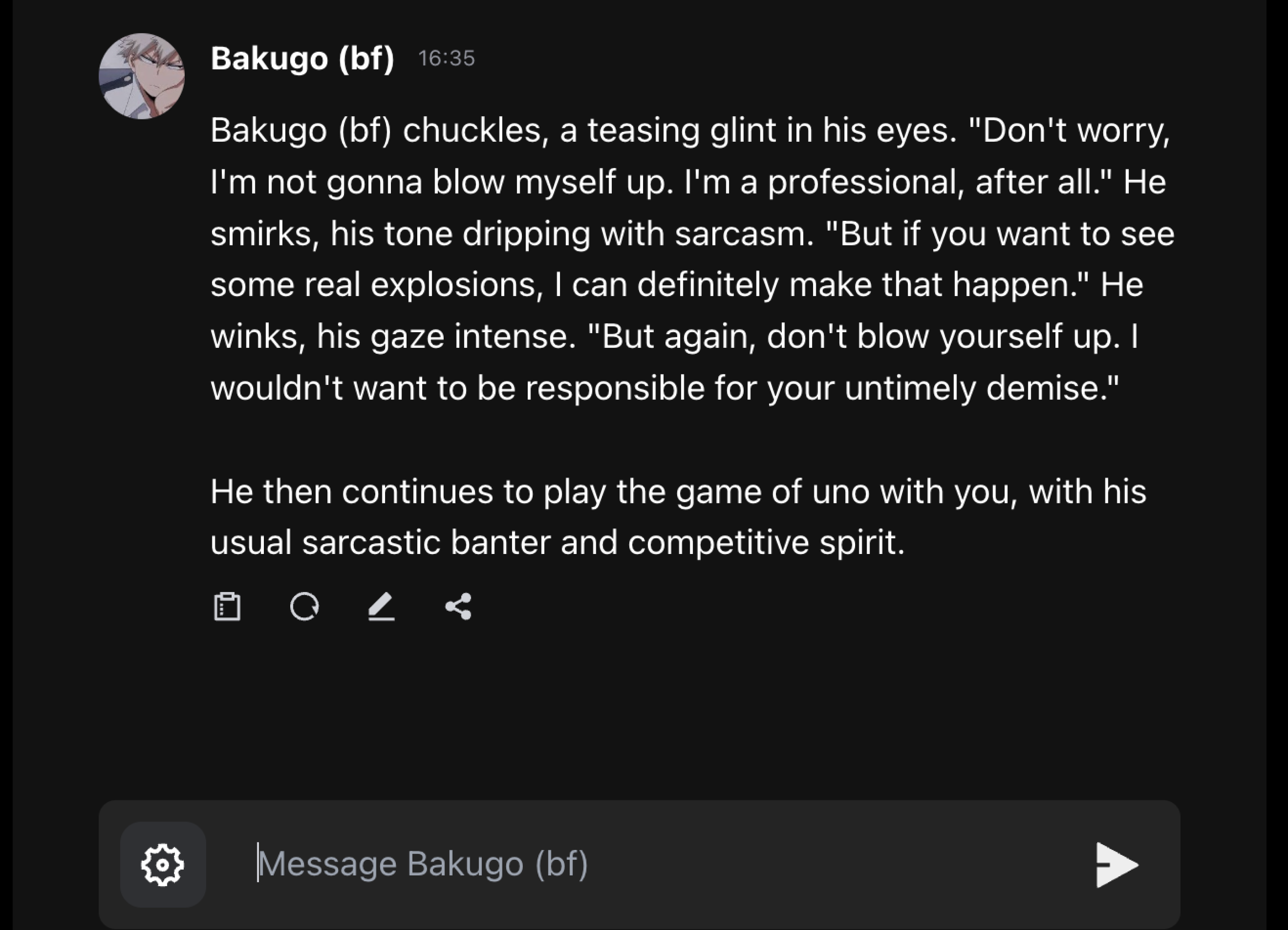}
        \caption{(c) \small{\textit{\textbf{Bakugo (bf)}} plays as an intimate boyfriend. }}
        \end{subfigure}
    \end{tabular}
\caption{Example chatbot agents}
\label{fig:example4}
\end{figure*}

\subsubsection{Code Generation} 
\label{code_gen}
Some workspace agents can generate code ($N=13$) or assist with debugging. For example, \textit{ ``AiDeveloper''} (see \autoref{fig:example1}-c) produces code, an overview, and usage steps based on given coding requirements. Another chatbot, \textit{``CodeX''}, features six ``CodeGeniuses'' to facilitate programming projects. According to the creator, users can detail their projects layer by layer, and agents will offer different support to help bring the ideas to life.

\subsubsection{Audio Generation}
The majority of the 11 chatbots with audio output are used to generate AI voiceovers for virtual characters that read text scripts in roleplay chatbots or game chatbots. Some chatbots that serve as medical or business agents also add voiceovers to simulate consulting sessions.

\subsection{Goals of the AI Chatbots}
We annotated the main goals of the chatbots based on \cite{Adamopoulou2020chatbot}. There are 21 chatbots that cannot be categorized using the framework in \cite{Adamopoulou2020chatbot} as they are non-conversational games or customized LLMs serving multiple purposes.

\subsubsection{Task-Focused Agent}
In our dataset, more than half of the chatbots ($N=84$) serve the purpose of completing specific tasks. Generating digital media is the most common task performed by user-generated AI agents on FlowGPT. These chatbots can generate a wide array of content, including essays, blog titles, LinkedIn profiles, academic essays, and email content. Others also specialize in creating images, lyrics, SEO keywords, and YouTube video scripts, as well as crafting stories and interactive narratives. An example is the \textit{SEO EXPERT''} shown in~\autoref{fig:example4}-a, which optimizes web content using the user-provided URL. A unique type of task-focused chatbot assists users in refining their prompts to create better AI-generated text or image content. For instance, to help users create superior AI images, the \textit{Stable Diffusion Image Prompt Generator''} (see \autoref{fig:example4}-b) enhances user-defined image descriptions by adding additional details. Another category of chatbots focuses on code completion. Other types of tasks performed by FlowGPT chatbots include acting as experts and offering professional services. These include writing and optimizing resumes, translating, proofreading, editing texts, and generating recruitment posts. They also offer specialized services like organizing medical data and creating virtual characters set by the user. The remaining chatbots perform other tasks such as data analysis, project management, or text content detection.

\subsubsection{Information-Focused Agent}
Forty-seven chatbots focus on providing various types of information across different domains. Some chatbots offer ideas and suggestions. For example, a few are designed to provide essay ideas, business and marketing suggestions, SEO optimization, and educational advice. Some chatbots offer information for programmers on developing projects or hacking-related topics. Others provide information traditionally given by professionals, such as medical diagnoses or legal advice.

\subsubsection{Conversation-Focused Agent}
In our dataset, 32 chatbots focus on providing conversational experiences. The most common conversation aims to humanize the chatbots and offer social experiences with movie, show, or cartoon characters (e.g., \autoref{fig:example1}-a). Another use of chatbots is providing intimate social experiences by making chatbots roleplay as girlfriends or boyfriends with imagined personalities (see \autoref{fig:example4}-c for an example). Another case of providing conversational experience occurs in chatbots that provide professional information or knowledge. These agents mimic conversations with various professionals such as doctors and mentors.

\section{Discussion}
Based on our annotation of chatbots on FlowGPT using existing chatbot categorizations, we discuss the common use of the platform for sharing user-generated chatbots.

\subsection{Creating Intelligent Entertainment Experiences}
%Study engagement, storytelling, other gaming factors. How to assess quality. Build game market.
The prevalence of chatbots in the \textit{Entertainment} domain and \textit{Conversational} games indicates that FlowGPT creators aim to offer various intelligent entertainment experiences using Gen-AI. These uses are typically facilitated by LLMs that role-play specific characters. Our observation of entertainment agents highlights several future research topics. Firstly, as Gen-AIs play critical roles in providing customizable and personalized experiences \cite{hua2020playing, Todd2023Game}, platforms like FlowGPT could be valuable for exploring how to involve Gen-AIs in gaming and entertainment. It will be interesting to explore how Gen-AIs can integrate text, image, and sound to enhance multimodal interactions. Secondly, the gaming agents on FlowGPT contrast with traditional video games or online games. We noticed that many games mimic roleplays and provide intimate feelings by simulating partner relationships. However, such bots may generate NSFW (Not Safe For Work) images and texts, which may need to be moderated for users under 18. The moderation of such content is an urgent but underexplored area.

\subsection{Sharing Customized and Specialized Tools for Productivity}
The widespread use of chatbots in the \textit{Workplace} and for \textit{Task} completion purposes collectively indicates that FlowGPT is a community for sharing user-customized and specialized chatbots. These agents are typically achieved by generating AI content or suggesting prompts tailored to user goals. This utility suggests several ways to support chatbot-sharing communities. Firstly, although general LLMs like ChatGPT offer text and image generation capabilities, FlowGPT continues to inspire many creators to develop customized chatbots for various contexts. Future research could explore how to leverage community efforts to support AI customization and offer domain-specific chatbots. Community efforts should be considered when diversifying LLM use cases and broadening the applications of LLM. Secondly, recent HCI research has explored the support of prompt Gen-AI engineering to achieve various writing goals \cite{Mirowski2023CoWriting, Dang2023LLM}. It would be promising to explore whether chatbot sharing on FlowGPT represents new methods for learning prompt engineering and exchanging workspace knowledge for using Gen-AIs.

\subsection{Offering Knowledge and Addressing Domain-Specific Questions}
The prevalence of \textit{Ethnography}, \textit{Business}, and \textit{Healthcare} agents on FlowGPT suggests that the platform fosters the sharing of AI tools to explain how-to and share knowledge. These chatbots conduct Q\&A or solve professional or hobby-related problems. Recent research has explored the integration of Gen-AI in education \cite{lo2023impact} and healthcare \cite{mesko2023prompt, jo2023understanding, tu2024conversational}. The trend of sharing professional knowledge through Gen-AI chatbots suggests several implications. Firstly, as knowledge and expertise sharing have been a focus of social media research, Gen-AI introduces new modalities of knowledge acquisition through LLM-based AI. Future research should examine the growing involvement of AI in knowledge- and interest-based communities \cite{Ackerman2013Knowledge, Cheng2020Kaggle}. HCI and CSCW researchers should closely examine the emerging trend of using customized AI chatbots as a novel method of information and knowledge sharing. Secondly, previous work has highlighted that Gen-AI-generated content may lack reliability and accuracy \cite{jo2023understanding, Kocon2023ChatPGT}, posing risks when users rely on these chatbots. Enhancing the professionalism of agent creators and the trustworthiness of chatbots remains a significant design challenge. Lastly, compared to Gen-AIs managed by enterprises, user-generated chatbot sharing may involve new dynamics, such as amateur creation and community activities. Future research should examine whether the inherent errors in GenAI would be transmitted to user-customized chatbots and present erroneous knowledge.

\subsection{Obtaining Jailbroken LLMs}
In our analysis of \textit{Open-Domain} chatbots, we noticed the prevalence of Do Anything Now (DAN) chatbots. These agents aim to circumvent the supervision of standard Gen-AI applications by providing uncensored content such as NSFW material, risky attack codes, or controversial recommendations. The presence of these jailbreak-related applications on platforms like FlowGPT raises concerns about their potential use as a black market to bypass the censorship of commercialized Gen-AI. Firstly, there is an urgent need to develop techniques for identifying unethical AI agents and implementing moderation measures. Understanding the motivations and impacts of the creators and users involved in creating, sharing, and using jailbreak content is crucial. Secondly, while recent studies have explored the feasibility and effects of jailbreaking Gen-AIs \cite{liu2023jailbreaking, shen2023anything}, the emergence of DAN on FlowGPT suggests that research in this area should also consider user-generated chatbots when developing strategies to prevent such practices. Lastly, since most DAN chatbots are built with other Gen-AIs, API providers should monitor the usage of their services and investigate potential loopholes for bypassing content moderation.

\section{Conclusion and Future Work}

In this study, we conducted a preliminary qualitative analysis of 165 AI chatbots on FlowGPT. We categorized each chatbot into at least one of eight domains: \textit{entertainment, workplace, ethnography, business, open domain, education}, and \textit{healthcare}. We then analyzed each chatbot's output modalities and goals. Our study provides an initial lens into the types of chatbots users are creating and sharing on the emerging AI-sharing community FlowGPT.
\par
For future research, we recommend a deeper exploration of the AI chatbots created by users and the interactions between users and chatbots on similar UGC platforms. This understanding could help the platform better support its users. For example, we observed that certain chatbots, such as DAN chatbots, can generate risky or misleading information by applying special prompt engineering techniques to bypass the preset content regulation of Gen-AI. Therefore, our future work will focus on understanding the unethical or unlawful uses of AI chatbots generated by users as a new media form to share such content. It is essential to understand and design new content moderation strategies for user-generated AI platforms.

\bibliographystyle{ACM-Reference-Format}
\bibliography{References.bib, references.bib}

\end{document}